\documentclass[aps,prl,twocolumn,floatfix,superscriptaddress]{revtex4-1}
\usepackage[utf8] {inputenc}
\usepackage[english] {babel}
\usepackage{graphicx}
\usepackage{dcolumn}
\usepackage{bm}
\usepackage{mathtools}
\usepackage{float}
\usepackage{multirow}
\usepackage{amsmath, amsfonts, amssymb}
\usepackage{physics}

\usepackage{lmodern}
\usepackage{textcomp}
\usepackage{t1enc}

\usepackage{graphicx}
\usepackage{setspace}

\usepackage[toc, page]{appendix}
\usepackage{standalone}

\usepackage{verbatim}
\usepackage{xcolor}

\usepackage{natbib}

\begin{document}


\title{\textit{Ab initio} determination of pseudospin for paramagnetic defects in SiC}

\author{Andr\'as Cs\'or\'e}
\affiliation{%
 Department of Atomic Physics, Budapest University of Technology and Economics, Budafoki \'ut 8., H-1111, Budapest, Hungary}

\author{Adam Gali}
\affiliation{%
 Wigner Research Centre for Physics, PO. Box 49, Budapest H-1525, Hungary}
 \affiliation{%
 Department of Atomic Physics, Budapest University of Technology and Economics, Budafoki \'ut 8., H-1111, Budapest, Hungary}

\date{\today}

\begin{abstract}
Paramagnetic point defects in solids may exhibit a rich set of interesting and not yet fully resolved physics. In particular, character of wavefunctions and electron-phonon coupling in these defects may highly influence their interaction with external magnetic fields. Complex interplay between the electronic orbitals, phonons and electron spin determines the effective pseudospin of the system that we demonstrate on vanadium and molybdenum defects in hexagonal silicon carbide (SiC) by means of \textit{ab initio} calculations. In this Letter, we find a giant anisotropy in the $g$-tensor of these defects with Kramers doublet spin ground state, resulting in reduced and vanishing interaction with the magnetic field in parallel and transverse directions, respectively. The consequences of our finding in the application of these defects for quantum information processing are briefly discussed. 
\end{abstract}

\maketitle


Point defects may introduce levels in the fundamental band gap of semiconductors or insulators that radically change the optical and magnetic properties of the host material. In particular, these point defects could be paramagnetic, i.e. the electron spin is greater than zero. The simplest case introducing paramagnetic electronic structure is the Kramers doublet (KD) electron spin state. In particular, neutral vanadium (V)~\cite{KaufmannPRB1997, KunzerPRB1993, SpindlbergerPRA2019} and singly positive charge state of molybdenum (Mo) point defects~\cite{CsoreMSF2016, GallstromMSF2009, BosmaNPJ2018} in hexagonal SiC are identifed as KD systems.  Recently, highly anisotropic interaction with magnetic field governed by the corresponding $g$-tensors has been observed or tentatively proposed for these KD systems~\cite{BosmaNPJ2018, BaurPSSA1997}, however the underlying physics has not been understood so far. Here we note that, although parallel component ($g_\parallel$) of $g$-tensor was thoroughly studied for V in 6H SiC by invoking crystal field theory, the emerging strong electron-phonon coupling is not included~\cite{KaufmannMSF1995, DornenMSF1992, MaierMSF1992, ReinkeIOP1993, KunzerPRB1993} or considered as minor effect~\cite{KaufmannPRB1997}. The strong anisotropy in $g$-tensor can be described by an effective Hamiltonian and pseudospins in which parameters cannot be predicted by applying simple models, and rather fitting procedure to known experimental data is applied. In order to identify and understand these spin-related phenomena and correct interpretation of experimental data, there is an urgent need to determine these spin-Hamiltonian parameters from first principles calculations.  

In this Letter, we demonstrate that \textit{ab initio} calculations can reproduce the anisotropy of the observed $g$-tensors for V and Mo defects in hexagonal SiC. We show that both electron-phonon coupling manifested as dynamic Jahn-Teller (DJT) effect~\cite{Bersuker2006} and the character of the wavefunction will determine the pseudospin of the system, i.e.\ its interaction with the external magnetic field. We show that the complex interplay of electronic orbitals, phonons and spins results in modified parallel ($g_\parallel$) component with respect to the free electron $g$ factor ($g_0 = 2.003$) and vanishing transverse ($g_\perp$) component. We discuss the relevance of our results in the light of realization of telecom wavelength solid state qubits. 

\begin{figure} [b]
\centering
\includegraphics[width=0.45\textwidth]{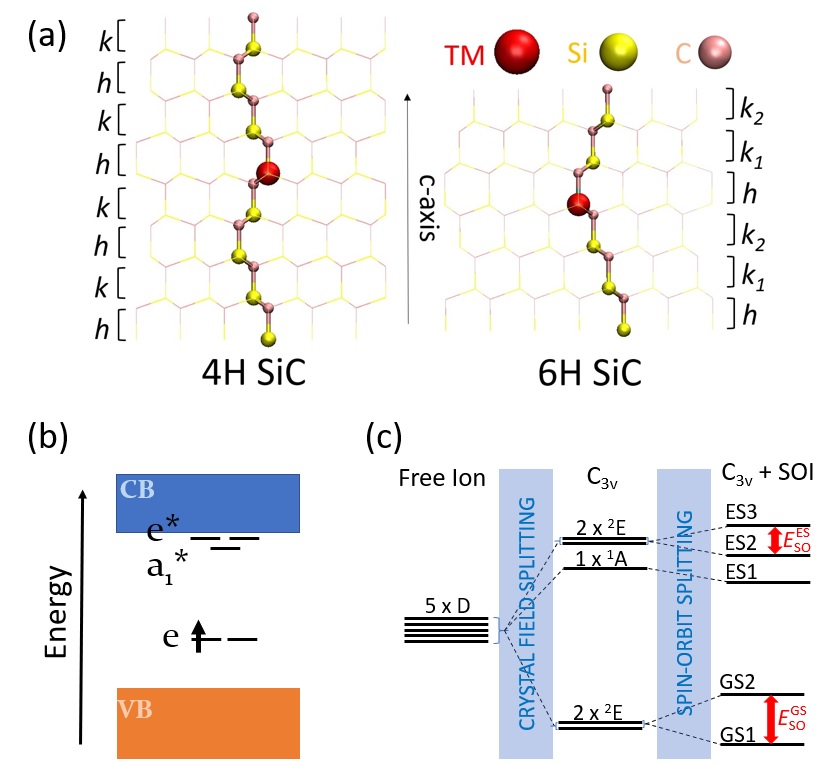}
\caption{\small (a) Employed 4H (left) and 6H (right) SiC supercells in orthographic view embedding a TM atom at a Si(\textit{h}) site. Crystal axis ($c$-axis), bilayer structures and labeling of atoms are indicated. (b) One-electron structure and (c) splittings in $d$-orbitals of TM$_\text{Si}(h)$ defects in both 4H and 6H SiC. In the level structure of TM$_\text{Si}(k)$ defects order of $a_1^*$ and $e^*$ levels are reversed and hence ${}^2A_1$ and the upper ${}^2E$ levels are also swapped. Energy of SO splittings in the ground and excited states are denoted by $E_\text{SO}^\text{GS}$ and $E_\text{SO}^\text{ES}$, respectively.}
\vspace{-20 pt}
\label{fig:eldefstruct1}
\end{figure}

Both transition metal (TM) atoms substitute a Si atom in the SiC lattice as found in earlier studies~\cite{GallstromMSF2009, IvadyPRL2011}.  However, lattice structures of 4H and 6H polytypes offer inequivalent lattice sites implying TM$_\text{Si}$ (TM = \{Mo$^+$,V\}) defects to form two configurations \--- a hexagonal ($h$) and a quasicubic ($k$) one \--- in 4H, and three configurations \--- a hexagonal ($h$) and two quasicubic ($k_1, k_2$) ones \--- in 6H SiC [cf.~Fig.~\ref{fig:eldefstruct1}(a)] all exhibiting C$_\text{3v}$ symmetry. For V$_\text{Si}$ all configurations have been observed in both 4H~\cite{SpindlbergerPRA2019} and 6H SiC~\cite{KaufmannPRB1997, KunzerPRB1993, KaufmannMSF1995, DornenMSF1992, MaierMSF1992, ReinkeIOP1993}. Recently, we have conclusively identified V$_\text{Si}$ configurations in 4H SiC~\cite{SpindlbergerPRA2019}. Identification of V$_\text{Si}$ configurations in 6H SiC is possible via the corresponding spin-orbit (SO) splittings and $g$-factors that is provided in this Letter. On the other hand, only a single signal has been detected for Mo$^+_\text{Si}$ in both hexagonal SiC polytypes as reported in recent PL studies~\cite{GallstromMSF2009, BosmaNPJ2018}. Since there is uncertainty with respect to the Mo$^+_\text{Si}$ configuration being responsible for the single PL signal, we investigate both defect models in 4H SiC, i.e. Mo$^+_\text{Si}(h)$ and also Mo$^+_\text{Si}(k)$. Our detailed results on Mo$^+_\text{Si}$ defect models are provided in the Supplementary Information~\cite{Supplementary}.

TM$_\text{Si}$ defects were embedded in a 576-atom 4H supercell and a 432-atom 6H supercell. For sampling the Brillouin-zone we used $\Gamma$-point which ensures the correct degeneracy of orbitals in C$_\text{3v}$ symmetry. Plane wave expansion of Kohn-Sham wavefunctions with a cutoff of 420~eV was applied as a natural choice for supercell-method. Relaxed geometries were achieved by minimizing the total energy with respect to the coordinates of the ions with fixed lattice constants of the perfect crystal where the corresponding quantum mechanical forces are prescribed to fall below 0.01~eV/$\AA$. We treated the core electrons within the framework of Projector Augmented Wave (PAW) method~\cite{BlochlPRB1994, Supplementary} as implemented in the \textsc{VASP} code~\cite{KressePRB1996}. In order to compute the spin-orbit (SO) splitting in the ground state we employed noncollinear approach~\cite{SteinerPRB2016} with fixed spin quantization axis along the crystal axis ($c$-axis), where the geometry was fixed in C$_\text{3v}$ configurations as obtained from spinpolarized calculations. The total energy was converged to 10$^{-8}$~eV in SOC calculations. From SO calculations orbitally reduced angular momentum (discussed later on) for each KS orbital can be directly read out~\cite{SteinerPRB2016}. We employed density functional theory (DFT) to calculate the electronic structure within the hybrid-DFT~+~V$_w$ scheme introduced by Iv\'ady~\textit{et~al.}~\cite{HeydJCP2003, IvadyPRB2014}. For the corresponding $w$ values in the ground state we found $w_\text{Mo} \approx 0$~eV for Mo$_\text{Si}^+$~\cite{CsoreMSF2016} and $w_\text{V} = 2.2$~eV for V$_\text{Si}$~\cite{SpindlbergerPRA2019}. We briefly describe this technique in Ref.~\cite{Supplementary}.

Both TM$_\text{Si}$ defects introduce spin doublet ($S = \frac{1}{2}$), i.e. KD ground state is formed by a single electron residing on a degenerate in-gap $e$ level. In addition, higher-energy empty $a_1$ and $e$ levels also occur in the band gap as plotted in Fig.~\ref{fig:eldefstruct1}(b), however their energy order is site dependent, i.e. $a_1(0)e(0)$ for TM$_\text{Si}(h)$ and $e(0)a_1(0)$ for TM$_\text{Si}(k)$ as already reported in Ref.~\onlinecite{SpindlbergerPRA2019}. 
As a result the ground state all-electron wavefunction transforms as $^2E$. However, the lower degenerate $e$ level is split by the SO coupling resulting in two SO sublevels in the ground state denoted by GS1 and GS2 in energy order [cf.\ Fig.~\ref{fig:eldefstruct1}(c)]. All in-gap one-electron levels exhibit closely atomic-like $d$-orbital character, thus the symmetry of the GS1-2 KD wavefunctions may be determined by the atomic KD states for each TM$_\text{Si}$ as constructed from $d$-orbitals (see Table~\ref{tab:kramerstab})~\cite{BosmaNPJ2018}. In particular, states can be established as proper linear combinations of $\Psi_{1-4}$ (see~Table~\ref{tab:kramerstab}) providing that GS1-2 transform either as $E_{\frac{1}{2}}$ or $E_{\frac{3}{2}}$ representations in the C$_\text{3v}$ double group notation. Here we note that, a close inspection on the local environment of the different sites implies that the $h$ site exhibits the weaker C$_\text{3v}$ (and thus stronger T$_\text{d}$) character than the $k$ or $k_1/k_2$ sites, while $k$ and $k_2$ show the strongest C$_\text{3v}$ nature. Consequently, SO splitting is expected to be the lowest for TM$_\text{Si}(h)$ defects, since in T$_\text{d}$ symmetry it is entirely quenched in the first order based on group theory considerations.

\begin{table}[t]
\begin{ruledtabular}
\caption{Kramers doublets formed by $d$-orbitals and the corresponding single and double group irreducible representations under C$_\text{3v}$ symmetry. We give widespread notations for double group irreducible representations (irreps) and also the corresponding $m_j$ values.}
\label{tab:kramerstab}
\begin{tabular}{@{}ccccc@{}}
\multirow{2}{*}{labels}
& \multirow{2}{*}{orbitals}
& \multicolumn{2}{c} {irreps.}
& \multirow{2}{*}{$m_j$}\\
& & single & double \\
\hline
$\Psi_1$ & $\ket{{d_{+2}, +\frac{1}{2}}}; \ket{{d_{-2}, -\frac{1}{2}}}$ & $^2E$ & $E_{\frac{1}{2}}$ ($\Gamma_4$) & $\pm \frac{5}{2}$ \\
$\Psi_2$ & $\ket{{d_{+2}, -\frac{1}{2}}}; \ket{{d_{-2}, +\frac{1}{2}}}$ & $^2E$ & $E_{\frac{3}{2}}$ ($\Gamma_{5,6}$) & $\pm \frac{3}{2}$\\
$\Psi_3$ & $\ket{{d_{+1}, +\frac{1}{2}}}; \ket{{d_{-1}, -\frac{1}{2}}}$ & $^2E$ & $E_{\frac{3}{2}}$ ($\Gamma_{5,6}$)& $\pm \frac{3}{2}$\\
$\Psi_4$ & $\ket{{d_{+1}, -\frac{1}{2}}}; \ket{{d_{-1}, +\frac{1}{2}}}$ & $^2E$ & $E_{\frac{1}{2}}$ ($\Gamma_4$)& $\pm \frac{1}{2}$\\
$\Psi_5$ & $\ket{{d_{0}, +\frac{1}{2}}}; \ket{{d_{0}, -\frac{1}{2}}}$ & $^2A_1$ & $E_{\frac{1}{2}}$ ($\Gamma_4$)& $\pm \frac{1}{2}$\\
\end{tabular}
\end{ruledtabular}
\end{table}

In this work, we aim to reveal the pivotal role of shape of the wavefunction and the interaction of the described electronic structure with the SiC phonon bath resulting in the highly anisotropic $g$-tensors. To this end, we derive the parallel and transverse elements of the diagonalized $g$-tensor ($\mathbf{g}$) from the following spin Hamiltonian including the SO coupling ($\hat{H}_\text{SO}$), the Zeeman effect ($\hat{H}_\text{Zee}$) upon external magnetic field and the hyperfine (HF) interaction ($\hat{H}_\text{HF}$) as $\hat{H}_\text{eff} = \hat{H}_\text{SO} + \hat{H}_\text{Zee} + \hat{H}_\text{HF}$, where
\begin{align}
\hat{H}_\text{SO} = -p\lambda_0\hat{L_z}\hat{S_z}, && \hat{H}_\text{Zee} = - \mathbf{B}\hat{\pmb{\mu}}, && \hat{H}_\text{HF} = \hat{\mathbf{I}}\mathbf{A}\hat{\mathbf{S}} \text{.}
\label{eq:SOZeehami}
\end{align}
In the expression of $\hat{H}_\text{SO}$ (Eq.~\ref{eq:SOZeehami}) $\hat{S}_z$ and $\hat{L}_z$ are the $z$-components of the spin and angular momentum operators, $\hat{\mathbf{S}}$ and $\hat{\mathbf{L}}$, respectively, $\lambda_0$ is the SO constant and $p$ stands for the so-called Ham reduction factor~\cite{HamPR1968, Bersuker2006} arising from the electron-phonon coupling. HF interaction is definitely active because of the presence of the impurity atom with non-zero nuclear spin. In the HF Hamiltonian $\hat{\mathbf{I}}$ is the nuclear spin operator and $\mathbf{A}$ is the HF tensor which can be separated into a Fermi-contact and a dipolar spin-spin interaction where the latter can rotate the electron spin because of the spin ladder operators ($S^+I^- + S^-I^+$). In the formula of $\hat{H}_\text{Zee}$ (Eq.~\ref{eq:SOZeehami}) $\mathbf{B}$ represents the external magnetic field and the dipole momentum operator ($\hat{\pmb{\mu}}$) can be expressed as
\begin{equation}
\hat{\pmb{\mu}} = - (\mu_\text{B}pr\hat{\mathbf{L}} + \mu_\text{B}g_0\hat{\mathbf{S}}) = - \mu_\text{B}\mathbf{g}\mathbf{\tilde{S}}\text{,}
\label{eq:muop}
\end{equation}
where the contributions of $\hat{\mathbf{L}}$ and $\hat{\mathbf{S}}$ are separated~\cite{ThieringPRX2018} and $\hat{\pmb{\mu}}$ is also expressed in the phenomenological pseudospin ($\mathbf{\tilde{S}}$) formalism~\cite{AbragamOxford1970}, where $\mathbf{\tilde{S}} = \frac{1}{2}$ for KD systems~\cite{ChibotaruJCP2012}. In Eq.~\ref{eq:muop} $\mu_\text{B}$ is the Bohr magneton, $g_0$ is the free electron spin $g$-factor and $r$ represents the Stevens orbital reduction factor~\cite{StevensProcRoySoc1953}. From Eq.~\ref{eq:muop} parallel ($g_\parallel$) and transverse ($g_\perp$) components of the $g$-tensor ($\mathbf{g}$) can be expressed as
\begin{align}
\label{eq:gpar}
g_\parallel &= 2(g_0S_z + L_z^\text{eff}) = \frac{2 \mu_z}{\mu_B}\text{,}  \\
\label{eq:gperp}
g_\perp &= \frac{ \mu_+ + \mu_- + i(\mu_- - \mu_+)}{\mu_B}\text{,}
\end{align}
where we use expectation values of the ladder dipole moment operators ($\mu_\pm$) to express $g_\perp$ (Eq.~\ref{eq:gperp})~\cite{BosmaNPJ2018}. In Eq.~\ref{eq:gpar} $S_z$ and $L_z^\text{eff}$ are expectation values of $\hat{S}_z$ and the effective angular momentum operator, $\hat{L}_z^\text{eff} = pr\hat{L}_z$, respectively. 

Foremost, we calculate $g_\parallel$ based on Eq.~\ref{eq:gpar} via obtaining $\hat{L}_z^\text{eff}$ arising from the double reduction of the atomic angular moment, $L_z$. Eigenvalue ($L_z^\text{o}$) of the orbitally reduced angular momentum operator $\hat{L}_z^\text{o} = r\hat{L}_z$ can be directly read out from \textit{first principles} calculations (see~Table~\ref{tab:JTSOC}). To determine the eigenvalue ($L_z^\text{eff}$) of the effective angular moment operator $\hat{L}_z^\text{eff} = p\hat{L}_z^\text{o}$, we account for the emerging strong electron-phonon coupling by calculating $p$. Finally, we determine $g_\perp$ according to Eq.~\ref{eq:gperp}.

\begin{table*}[t]
\begin{ruledtabular}
\caption{Parameters of the corresponding quadratic DJT APES ($E_\text{JT}$, $\delta_\text{JT}$) allowing the calculation of the effective phonon energy ($\hbar\omega$) and Ham reduction factor ($p$). Intrinsic (reduced) SO splitting energies ($\Delta E_\text{SOC}^\text{(red)}$) are also presented. Orbitally reduced (effective) angular momenta ($L_z^\text{o(eff)}$) for GS1 and GS2 and difference between them $L_z^\text{eff}$ values ($\Delta L_z^\text{eff}$) are also provided for V$_\text{Si}$ in 4H and 6H SiC.}
\label{tab:JTSOC}
\begin{tabular}{@{}cccccccccccc@{}}
\multirow{2}{*}{Polytype}
& \multirow{2}{*}{Site}
& \multicolumn{1}{c}{$E_\text{JT}$}
& \multicolumn{1}{c}{$\delta_\text{JT}$}
& \multicolumn{1}{c}{$\hbar\omega$}
& \multirow{2}{*}{$p$}
& \multicolumn{1}{c}{$\Delta E_\text{SOC}$}
& \multicolumn{1}{c}{$\Delta E_\text{SOC}^\text{red}$}
& \multicolumn{1}{c}{$\Delta E_\text{SOC}^\text{exp}$}
& \multirow{2}{*}{$L_z^\text{o}$ (GS1, GS2) }
& \multirow{2}{*}{$L_z^\text{eff}$ (GS1, GS2)}
& \multirow{2}{*}{$\Delta L_z^\text{eff}$}\\
& & (meV) & (meV) & (meV) & & (GHz) & (GHz) & (GHz) & & & \\
\hline
\multirow{2}{*}{4H} & $h$ & 9.4 & 5.6 & 60.19 & 0.63 & 9.91 & 6.29 & 43 ($\beta$) & -0.022, -0.013 & -0.014,-0.009 & 0.005 \\
 & $k$ & 13.1 & 7.1 & 49.81 & 0.60 &  819.21 & 490.37  & 529 ($\alpha$) & -0.125,0.094 & -0.069,0.059 &  0.128 \\
\cline{2-12}
\multirow{3}{*}{6H} & $h$ & 9.3 & 5.8 & 47.19 & 0.57 & 24.18 & 13.78 & 16 ($\gamma$) & -0.012,-0.018 & -0.007,-0.010 & 0.003\\
 &$k_1$ & 11.4 & 5.8 & 49.43 & 0.55 & 82.94 & 45.62 & 25 ($\beta$) & -0.015,-0.016 & -0.008,-0.009 & 0.001 \\
 &$k_2$ & 11.9 & 6.1 & 65.91 & 0.61 & 808.58 & 493.23 & 524 ($\alpha$) & -0.117,0.086 & -0.071,0.052 & 0.123\\
\end{tabular}
\end{ruledtabular}
\end{table*}

Ground state electronic structure introduced by the TM$_\text{Si}$ defects [cf.\ Fig.~\ref{fig:eldefstruct1}(b)], i.e.~the half-filled orbitally degenerate $e$ level may split by coupling with $e$ phonon modes as manifestation of $E \otimes e$ JT effect~\cite{Bersuker2006, HamPR1968}. Experimental results imply no symmetry reduction thus dynamic JT (DJT) is expected for these systems~\cite{SpindlbergerPRA2019, BosmaNPJ2018}. To estimate the magnitude of DJT effect originating from the electron-phonon coupling we calculated the trivial points of the quadratic DJT adiabatic potential energy surface (APES)~\cite{Bersuker2006}, i.e. the three minima (C$_\text{1h}$), the three barrier (C$_\text{1h}$) and the high symmetry (C$_\text{3v}$) points. Energy separation between the C$_\text{3v}$ and the three minima is the JT energy ($E_\text{JT}$), while barrier points are separated by the barrier energy ($\delta_\text{JT}$). A general quadratic DJT APES is shown in Fig.~I in Ref.~\cite{Supplementary}.
As a result angular momentum might be severely reduced by the persisting DJT effect known as Ham effect~\cite{HamPR1968, Bersuker2006} resulting in the reduction of the spin-orbit coupling (SOC) and the $g$-tensor elements. Reduction can be expressed as $p\hat{\mathbf{L}}$, where $p$ is the already introduced Ham reduction factor.
For calculation of $p$ corresponding APES have to be determined as described by the Hamiltonian~\cite{Bersuker2006} of
\begin{equation}
\begin{aligned}
\hat{H}_\text{DJT} &= \hbar\omega(a_x^{\dag}a_x + a_y^{\dag}a_y + 1) + F(x\sigma_z + y\sigma_x)\\  &+ G[(x^2-y^2)\sigma_z + 2xy\sigma_x],
\end{aligned}
\label{eq:HDJT}
\end{equation}
where $a_{x/y}^{(\dag)}$ represent annihilation (creation) operators of two-dimensional $e$ modes vibrating in the $xy$ plane and  electrons are represented by the Pauli matrices $\sigma_x$ and $\sigma_z$. In Eq.~\ref{eq:HDJT}, $\hbar\omega$ stands for the effective energy of the $e$ modes, while $F=\sqrt{E_\text{JT} \cdot 2\hbar\omega}$ and $G =\delta_\text{JT}\hbar\omega / 2E_\text{JT}$ govern the linear and quadratic nature of the APES, respectively. In this way all parameters in DJT Hamiltonian can be directly readout from the corresponding APES enabling the numerical solution of Eq.~\ref{eq:HDJT} and thus the determination of the polaronic wavefunctions that can be expanded in complex basis as
\begin{equation}
\ket{\Psi_\pm} = \sum_{nm} (c_{nm}\ket{E_\pm} \otimes \ket{n,m} + d_{nm} \ket{E_\mp} \otimes \ket{n,m})\text{.}
\label{eq:polwav}
\end{equation}
In Eq.~\ref{eq:polwav} $E_\pm$ represents the complex components of the ${}^2E$ ground state, where the subscript denotes the corresponding $L_z = \pm1$ formed by two $d$-orbitals as $E_-$:\{-2,+1\} and $E_+$:\{+2,-1\}.  States of $E_\pm$ are mixed by the vibronic wavefunctions of $\ket{n,m}$, where $n + m \leq 4$ basis set provides convergent $\ket{\Psi_\pm}$. In this way mixing coefficients of $c_{nm}$ and $d_{nm}$ can be calculated that enables us to determine $p$ via the formula of
\begin{equation}
p = \sum_{nm} (c^2_{nm} - d^2_{nm})
\label{eq:pfact}
\end{equation}
as derived and implemented by Thiering \textit{et al.}~\cite{ThieringPRX2018, ThieringPRB2017}. We provide the expansion of polaronic wavefunctions in symmetry-adapted basis~\cite{ThieringPRB2017, MazeIOP2011} in Ref.~\cite{Supplementary}. The corresponding calculated values of $E_\text{JT}$, $\delta_\text{JT}$ and $p$ are listed in Table~\ref{tab:JTSOC}. We report the ground state polaronic wavefunctions expanded in symmetry adapted basis for all vanadium defect configurations in Ref.~\cite{Supplementary}.

We also report the corresponding \textit{intrinsic} and reduced ground state SO splittings ($E_\text{SOC}$) in Table~\ref{tab:JTSOC} as obtained by our calculations. Reduced SOC can be calculated as 
\begin{equation}
\begin{aligned}
\Delta E_\text{SOC}^\text{red} &= p \Delta E_\text{SOC} = p(E_\text{SOC}^\text{GS2} - E_\text{SOC}^\text{GS1})\\
 &= p (\lambda_0^\text{GS2} \langle \hat{L_z^\text{o}}\hat{S_z} \rangle^\text{GS2}-\lambda_0^\text{GS1} \langle \hat{L_z^\text{o}}\hat{S_z} \rangle^\text{GS1}),
\label{eq:ESOC}
\end{aligned}
\end{equation}
where $\Delta E_\text{SOC}^\text{(red)}$ stands for the intrinsic (reduced) SOC splitting of the ground state, $\lambda_z^\text{GS1-2}$ is the intrinsic SOC constant of the SO sublevels GS1-2 and $\langle \hat{L_z^\text{o}}\hat{S_z} \rangle^\text{GS1-2}$ represents the expectation value of the $\hat{L_z^\text{o}}\hat{S_z}$ product. In the calculations we employ $S_z = +\frac{1}{2}$ for both GS1 and GS2. Since the $d$-orbitals are well-localized on the V impurity and almost the entire SO splitting originates from the V atom one may deduce from the atomic SO splitting formula that $\Delta E_\text{SOC}^\text{red}~\sim~\Delta L_z^\text{eff}$, where $\Delta L_z^\text{eff}~=~p(L_z^\text{o,GS2}~-~L_z^\text{o,GS1})$~\cite{Supplementary}.

We list the calculated values of $E_\text{SOC}$ and $E_\text{SOC}^\text{red}$ along with the experimental values~\cite{WolfowiczSciAdv2020} in Table~\ref{tab:JTSOC}. Accordingly, both calculated and experimental values for V$_\text{Si}(k)$ in 4H SiC and for V$_\text{Si}(k_2)$ in 6H SiC are at least one order of magnitude higher than that for V$_\text{Si}(h)$ in 4H  and V$_\text{Si}(h/k_1)$ in 6H SiC. This significant difference may be explained via the calculated $\Delta L_z^\text{eff}$ values also listed in Table~\ref{tab:JTSOC} where similar trend occurs obeying the previosuly deduced linear relationship between $\Delta E_\text{SOC}^\text{red}$ and $\Delta L_z^\text{eff}$. Furthermore, SO splitting for V$_\text{Si}(k_1)$ is slightly larger than that of V$_\text{Si}(h)$ arising from the local symmetry exhibiting stronger C$_\text{3v}$ character for V$_\text{Si}(k_1)$ than V$_\text{Si}(h)$.

At this point we are ready to calculate $g_\parallel$ according to Eq.~\ref{eq:gpar}: values are included in Table~\ref{tab:gfactor}. We find good agreement between the trends of the calculated and that of the experimental $g_\parallel$ values. In particular, $g_\parallel$ for V$_\text{Si}(k)$ in 4H and for V$_\text{Si}(k_2)$ in 6H SiC are well-separated from those of V$_\text{Si}(h)$ in 4H and V$_\text{Si}(h/k_1)$ in 6H SiC, respectively, supporting our defect identification based on the SO splittings. On the other hand, for the Mo point defects we found that $g_\parallel$ of both GS1 and GS2 of Mo$_\text{Si}^+(h/k_1)$ in both polytypes is lower than $g_0 = 2.0023$. In contrast, for Mo$_\text{Si}^+(k/k_2)$ $g_\parallel$ of GS1 is lower, while that of GS2 is higher than $g_0$.

 For the calculation of $g_\perp$ (Eq.~\ref{eq:gperp}), we consider the ladder magnetic dipole operator, $\hat{\mu}_\pm$ that can couple state $\ket{m_j}$ to state $\ket{m_j \pm 1}$, where $m_j~=~m_l~+~m_s$. However, GS1 and GS2 transform as either $E_{1/2}$ (linear combination of $\Psi_1$ and $\Psi_4$) or $E_{3/2}$ (linear combination of $\Psi_2$ and $\Psi_3$) with the $m_j$ values given in Table~\ref{tab:kramerstab}. Consequently, $\hat{\mu}_\pm$ cannot couple neither $\Psi_1$ and $\Psi_4$, nor $\Psi_2$ and $\Psi_3$ therefore $g_\perp = 0$ (cf.\ Table~\ref{tab:gfactor}) in each case. Here we note that second order contributions to $g_\perp$ might occur by mixing $\Psi_2$ or $\Psi_3$ with $\Psi_5$, since $\bra{\Psi_5}\hat{\mu}_\pm\ket{\Psi_5} \neq 0$. This may occur vibronically or by the hyperfine interaction. Experimental ground state HF parameters are available in the literature which are around 15-70~MHz for Mo$_\text{Si}^+$ in 6H SiC~\cite{BaurPSSA1997}; 160-230~MHz for V$_\text{Si}(h)$ and 100-190~MHz for V$_\text{Si}(k)$~\cite{WolfowiczSciAdv2020} in 4H SiC. HF will mix the corresponding wavefunctions only in the second order, thus it is expected that the final $g_\perp$ factor will be at least two orders of magnitude smaller than that of $g_\parallel$.

Calculated SO splitting and $g_\parallel$ values agree well with the experimental ones giving the possibility of defect identification. Accordingly, we identify the $\alpha$ and $\beta$ centers~\cite{WolfowiczSciAdv2020, SpindlbergerPRA2019} as V$_\text{Si}(k)$ and V$_\text{Si}(h)$ in 4H SiC, respectively, supporting the earlier considerations reported in Ref.~\cite{SpindlbergerPRA2019}. Regarding 6H SiC, we identify the $\alpha$ center as V$_\text{Si}(k_2)$ while SO splittings and values for $g_\parallel$ of V$_\text{Si}(h)$ and V$_\text{Si}(k_1)$ are not well-separated~\cite{WolfowiczSciAdv2020} for immediate identification from the calculated $g$-constants and further considerations are needed. To this end, we calculated the corresponding zero-phonon lines (ZPL) that resulted in larger energy for $k_1$ than that for $h$ configuration~\cite{Supplementary}. As a result we associate the $\alpha$ signal with V$_\text{Si}(k/k_2)$ and the $\beta$ line with V$_\text{Si}(h/k_1)$ in 4H/6H SiC and the $\gamma$ center is identified as V$_\text{Si}(h)$ in 6H SiC.

\begin{table}[b]
\begin{ruledtabular}
\caption{Experimental and calculated values of $\mathbf{g}$-tensor elements ($g_\parallel, g_\perp$) for GS1 and GS2 of all possible V$_\text{Si}$ and Mo$_\text{Si}$} configurations in 4H and 6H SiC.
\label{tab:gfactor}
\begin{tabular}{@{}cccccc@{}}
\multirow{3}{*}{Defect}
& \multirow{3}{*}{Site}
& \multicolumn{2}{c}{Experiment}
& \multicolumn{2}{c}{Theory}\\
& & GS1 & GS2 & GS1 & GS2 \\
& (PL center) & $g_\parallel$,$g_\perp$ & $g_\parallel$,$g_\perp$ & $g_\parallel$,$g_\perp$ & $g_\parallel$,$g_\perp$ \\
\hline
\multirow{2}{*}{4H-V$_\text{Si}$} & $h (\beta) $ & 1.870,<1\footnotemark[1] & 2.035,<1\footnotemark[1] & 1.975,0 & 1.987,0\\
 & $k (\alpha)$ & 1.748,0\footnotemark[2] & 2.160,0\footnotemark[2] & 1.866,0 & 2.106,0\\
\cline{2-6}
\multirow{3}{*}{6H-V$_\text{Si}$} & $h (\gamma)$ & 1.933,<1\footnotemark[1] & 1.972,<1\footnotemark[1] & 1.989,0 & 1.983,0\\
 &$k_1 (\beta)$ & 1.95,-\footnotemark[1] & 2.00,-\footnotemark[1] & 1.987,0 & 1.985,0 \\ 
 &$k_2 (\alpha)$ & 1.749,0\footnotemark[2] & -,- & 1.860,0 & 2.108,0\\
\cline{2-6}
\multirow{2}{*}{4H-Mo$_\text{Si}$} & $h$ & \multirow{2}{*}{1.87,0.04\footnotemark[3]} & \multirow{2}{*}{-,-} & 1.976,0 & 1.990,0\\
 & $k$ & & & 1.915,0 & 2.063,0\\
\cline{2-6}
\multirow{3}{*}{6H-Mo$_\text{Si}$} & $h$ & \multirow{3}{*}{1.610,0\footnotemark[3]} & \multirow{3}{*}{-,-} & 1.980,0 & 1.979,0\\
 & $k_1$ & & & 1.985,0 & 1.994,0 \\ 
 & $k_2$ & & & 1.919,0 & 2.059,0\\
\end{tabular}
\end{ruledtabular}{
\footnotetext[1]{Ref.~\cite{WolfowiczSciAdv2020} }
\footnotetext[2]{Ref.~\cite{KaufmannPRB1997, BaurPSSA1997}}
\footnotetext[3]{Ref.~\cite{BosmaNPJ2018}}}
\end{table}

Our results have implications on quantum information processing based on solid state defect qubits. It has been proposed~\cite{BosmaNPJ2018, SpindlbergerPRA2019} that Mo and V dopants in SiC with optical transitions near or inside telecom wavelength bands make possible to integrate these solid state qubits to telecommunication technology. Indeed, all-optical identification and
coherent control of ensemble Mo center have been realized~\cite{BosmaNPJ2018}. Parallel to our study, vanadium defects have been isolated and coherent control of single spins have been demonstrated with showing all the ingredients required for
a highly efficient spin-photon interface~\cite{WolfowiczSciAdv2020}. Our study shows the nature of the ground state spin of these systems, namely, the order of spin levels and the origin of zero-field-splitting. This knowledge is crucial in optimizing the quantum optics protocols. Our results reveal the microscopic mechanism behind the phenomena of giant anisotropy in the interaction of the electron spin with the external stray magnetic fields which is only observable in the parallel component but minor in the transverse components.
 
In summary, we carried out hybrid-DFT calculations in order to reveal microscopic origin behind the highly anisotropic magnetic properties of KD systems as observed in experiments~\cite{BosmaNPJ2018}. To this end, we shed light on the ground state electronic structure and we calculated the corresponding SO splittings. We found that V$_\text{Si}(k)$ and V$_\text{Si}(k_2)$ exhibit one order of magnitude larger SO splittings than that of V$_\text{Si}(h)$ in 4H and V$_\text{Si}(h)$, V$_\text{Si}(k_1)$ in 6H SiC, respectively. This allowed us to identify the $\alpha$ centers in 4H and 6H SiC as V$_\text{Si}(k)$ and V$_\text{Si}(k_2)$, respectively, while the $\beta$ center in 4H SiC corresponds to V$_\text{Si}(h)$. From the spin Hamiltonian we derived $g_\perp$ and $g_\parallel$ (Eqs.~\ref{eq:gpar}~and~\ref{eq:gperp}) and found good agreement between the known experimental and calculated values for V$_\text{Si}$ supporting the significant role of electron-phonon coupling and character of the wavefunction in evolving the interaction of the electron spin with the magnetic field. 

The support from \'UNKP-19-3 New National Excellence Program of the Ministry of Human Capacities of Hungary is acknowledged by A.\ Cs. A.\ G.\ acknowledges the National Research, Development, and Innovation Office of Hungary grants No.~KKP129866 of the National Excellence Program of Quantum-coherent materials project, No.~127902 of the EU QuantERA Nanospin project, No.~2017-1.2.1-NKP-2017-00001 of the National Quantum Technology Program, and the Quantum Information National Laboratory supported by the Ministry of Innovation and Technology of Hungary, as well as the EU Commission for the H2020 Quantum Technology Flagship projects ASTERIQS (Grant No.~820394) and QuanTelCO (Grant No.~862721).


%

\end{document}